\documentclass{ledger}

\newcommand{\thefirstpagenum}[0]{1}
\newcommand{\thelastpagenum}[0]{5}
\newcommand{\theyear}[0]{201X}
\newcommand{\thevol}[0]{X}
\newcommand{\thedoi}[0]{DOI 10.5195/LEDGER.\theyear.X}
\newcommand{\ledgerpages}[0]{\thefirstpagenum-\thelastpagenum}

\hypersetup{pdfauthor={First Last}, pdftitle={Blockchain in a box: A portable blockchain network implementation on Raspberry Pi's}}

\title{Blockchain in a box: A portable blockchain network implementation on Raspberry Pi's}
\author{
Matija~Piškorec\thanks{Corresponding author (matija.piskorec@irb.hr)} $^{}$ \thanks{Rudjer Boskovic Institute, Zagreb, Croatia} $^{}$ \thanks{Blockchain and Distributed Ledger Technologies, University of Zurich, Switzerland} \and
Anton Ivashkevich\thanks{University of Zurich, Switzerland} \and
Said Haji Abukar\protect\footnotemark[4] \and
Lundrim Azemi\protect\footnotemark[4] \and
Md Rezuanul Haque\protect\footnotemark[4] \and
Mostafa~Chegenizadeh\protect\footnotemark[3] \and 
Claudio J. Tessone\protect\footnotemark[3] $^{~}$ \thanks{UZH Blockchain Center, University of Zurich, Switzelrand}
}

\pretitle{
  \centering \selectfont RESEARCH ARTICLE \par 
  \fontsize{24pt}{28pt}\selectfont} % Title is centered and at 24pt

\begin{document}

\maketitle

\thispagestyle{pagefirst}

\begin{abstract}
In this paper we describe a prototype of a blockchain-in-a-box system which allows users to easily bootstrap the whole Ethereum Proof-of-Work (PoW) network running on multiple Raspberry Pi nodes - an inexpensive modular computers. Users are able to orchestrate the whole blockchain network using a single web based interface, for example they are able to set the topology of the peer-to-peer (P2P) connections and control the initialization parameters. Each Raspberry Pi has a screen attached which visualizes current state of local blockchain, allowing users to easily visualize the consensus of the network in real time. We show how this platform can be used to perform experiments on consensus quality while using different P2P topologies. Similar experiments can be used for demonstration purposes in a workshop or other educational settings.  

% \todo{Approximately 150 words (200 words max), and currently it's 129 words!}

%AUTHOR: keywords are OK to show for Review article, will be hidden for publication
% \begin{keywords}
% \item Raspberry Pi
% \item Ethereum Proof-of-Work
% \item Consensus visualization
% \todo{Not more than two lines of keywords.}
% \end{keywords}
\end{abstract}

\section{Introduction}

% \todo{The Introduction should contain approximately 500 to 750 words (1000 words max) and should not use sub headings. Currently it's }

Initial ideal of a blockchain technology~\cite{Nakamoto2008} was for the regular users to be able to run their own blockchain nodes and to participate in the network as miners, potentially earning mining rewards for their contribution to securing the network. Having many node operators also ensures that blockchain network is sufficiently decentralized which contributes to the security of the network. However, due to rising requirements for computational resources it is rare that regular users are running their own blockchain nodes anymore. 

Building on our previous work on consensus visualization on Raspberry Pi's for the Bitcoin network~\cite{Ambrosini2022} as well our current efforts in building a similar system using Ethereum network~\cite{uzh-seminar-2023, uzh-thesis-2024, uzh-map-2024} in this paper we describe a fully functioning prototype of blockchain-in-a-box system capable of running an Ethereum Proof-of-Work (PoW) network on multiple Raspberry Pi's. We chose Ethereum PoW as a suitable blockchain because it's easy to run through well supported clients such as Go Ethereum (geth~\cite{geth}), onboarding of new users is easier than in PoS, and users are able to obtain the cryptocurrency through mining without relying on initial coin distribution from other nodes. Also, Ethereum is one of the most popular smart contract blockchains and its Ethereum Virtual Machine (EVM) is practically an industry standard for other smart contract blockchains, allowing users to deploy and interact with decentralized applications (DApps). We use Raspberry Pi nodes to demonstrate how easy it is to bootstrap the whole blockchain network even on a relatively modest hardware~\cite{Saingre2020,Fernando2019}.

Our blockchain-in-a-box prototype consists of a local server running on each Raspberry Pi that listens for the commands from a master server running on a separate regular consumer machine. The master server exposes a Web-based user interface through which users can orchestrate the initialization and operation of the blockchain network on the Raspberry Pi's. The prototype relies on a local network connectivity through a local WiFi router, so it is suitable to run in a workshop environment even without Internet connectivity. Each Raspberry Pi has a screen attached which shows the current state of its local blockchain, color coded in a visually simple manner so that it is easy to see whether the blockchain network as a whole is in consensus or not.

To demonstrate the usability of our prototype as an effective educational tool we run a series of experiments where we measure several consensus quality metrics with respect to the P2P topology of the nodes.

\section{Methods}

In this section we describe the components of our system along with the description of a typical experimental setup where blockchain is initialized from scratch for multiple Raspberry Pi nodes.

\subsection{Hardware setup}

\begin{figure}[ht]
  \centering
  \begin{minipage}[b]{0.49\textwidth}
    \includegraphics[width=\textwidth]{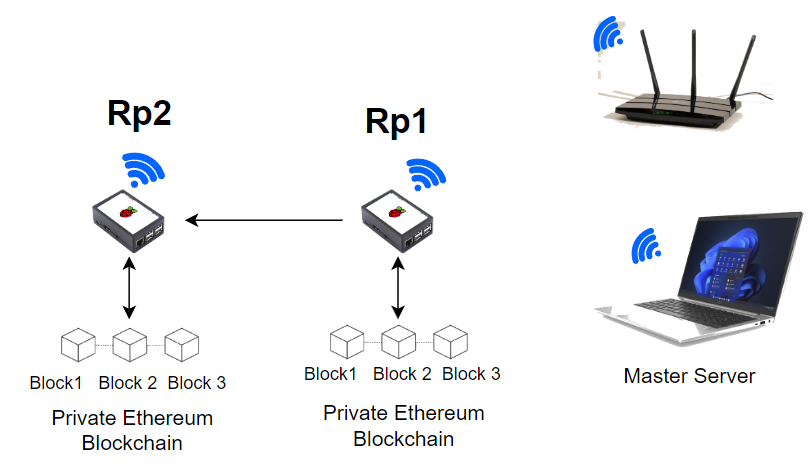}
  \end{minipage}
  \hfill
  \begin{minipage}[b]{0.49\textwidth}
    \includegraphics[width=\textwidth]{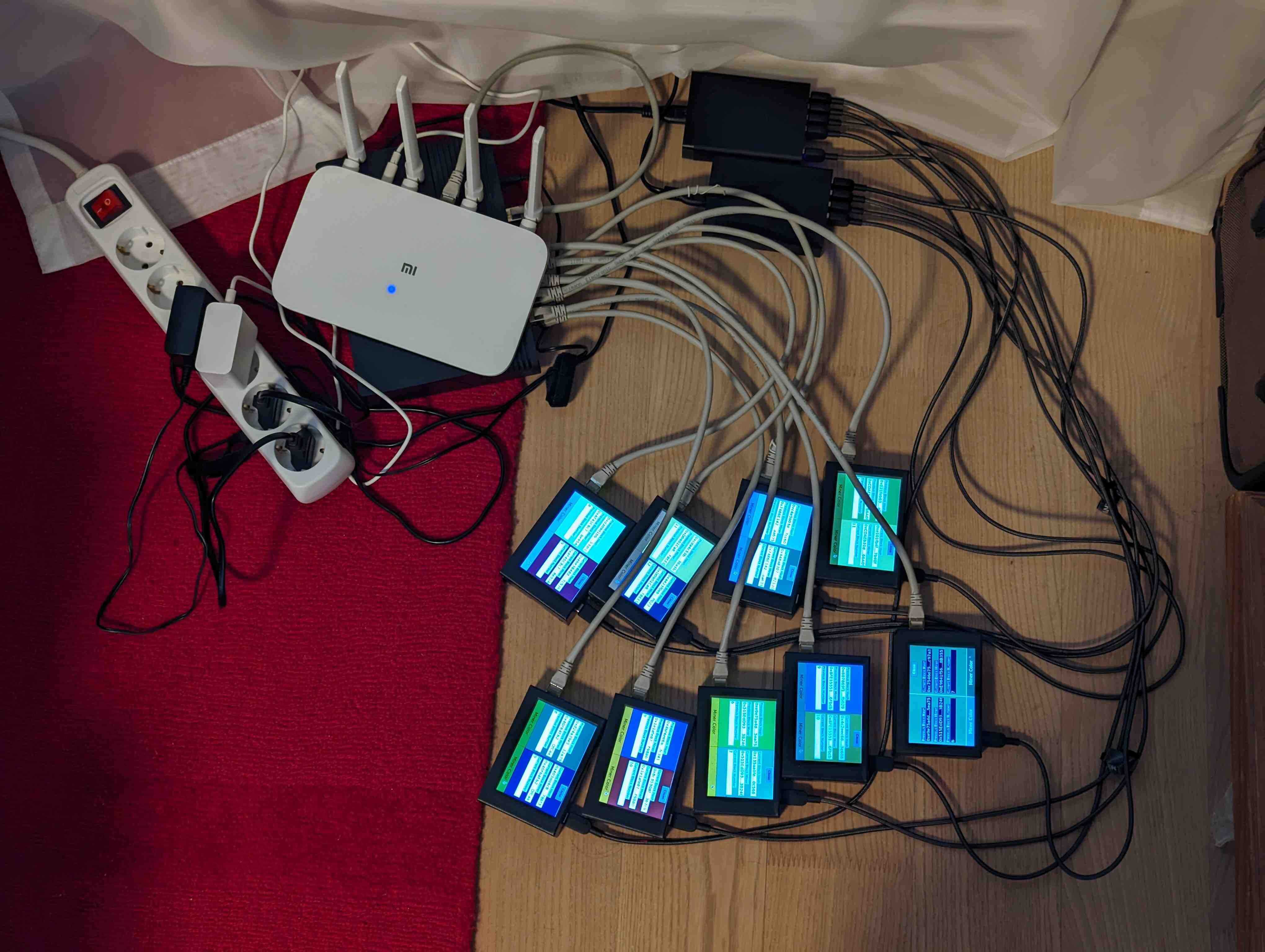}
  \end{minipage}
  \caption{An assembled hardware prototype (right) along with the hardware schematic (left).}
  \label{fig:experimental-setup}
\end{figure}

We choose Raspberry Pi computer to run our blockchain nodes. The reason for this its popularity, modular design, affordability and a wide open source support from a community of enthusiasts. In this work we use an 8Gb RAM version running 64-bit version of the RaspbianOS. Along with a basic computation module we use an 3.5 inch resistive Touch Screen and a metal case that encloses it. Rest of the hardware setup includes a Wi-Fi Router to provide a local network, a network switch used as the port extension in case of the wired setup, since router contains only 2 Ethernet ports, a multi-socket USB chargers to power multiple assembled processing units from the one power outlet, and a network and power cables for each assembled processing unit. Wi-Fi router and the Raspberry Pi's are configured so that each node gets a persistent IP address, which simplifies setting up the P2P topology between blockchain nodes. \cref{fig:experimental-setup} shows an assembled hardware prototype.

\vspace{0.25cm}
\vspace{0.25cm}

\subsection{Software setup}

\begin{figure}[ht]
  \centering
  \begin{minipage}[b]{0.49\textwidth}
    \includegraphics[width=\textwidth]{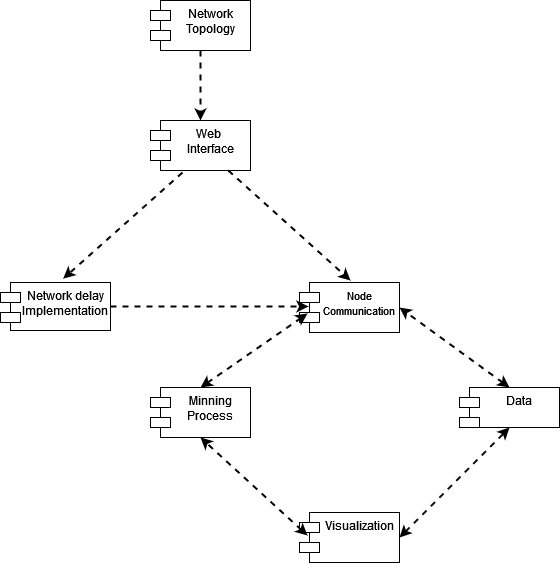}
  \end{minipage}
  \hfill
  \begin{minipage}[b]{0.49\textwidth}
    \includegraphics[width=\textwidth]{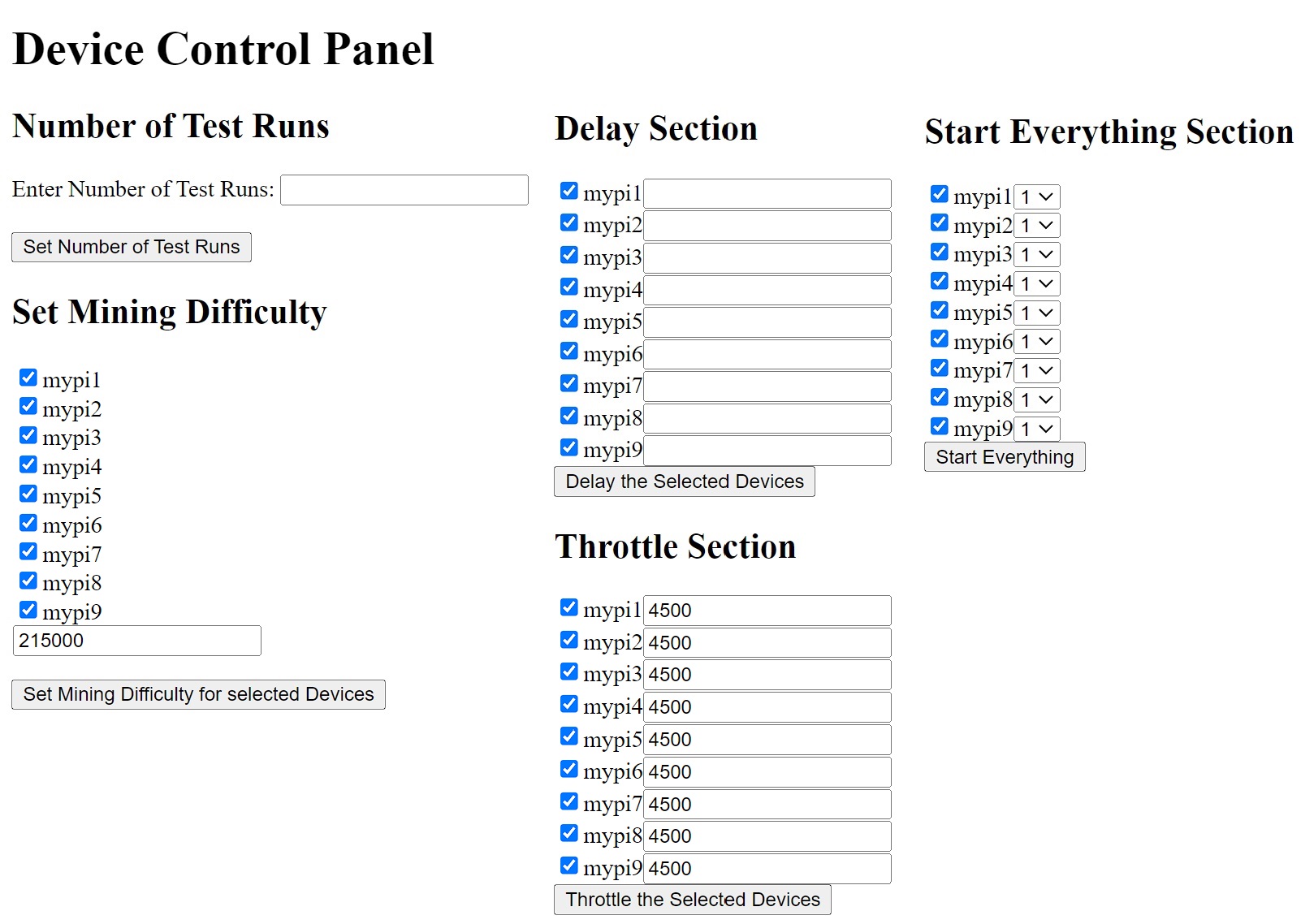}
  \end{minipage}
  \caption{A software setup of the prototype (left) with a master server web-based control panel allowing users to orchestrate multiple Raspberry Pi's from a single interface (right).}
  \label{fig:software-diagram}
\end{figure}

\cref{fig:software-diagram} shows the software setup of the prototype. Each RPi is running a local server which responds to the commands from the master server that is running on a separate machine (not necessarily a RPi). Users control the master server through a web-based control panel where experimental parameters can be adjusted. Each RPi runs a local server that listens for the commands coming from the master server and which is capable of initializing and managing a local blockchain node - in our case we use a Go Ethereum (geth) implementation of an Ethereum PoW execution client which is functioning both as a blockchain and a mining node. Local server is also responsible for logging local blockchain information in a format suitable for analysis, and managing a separate visualization component that displays current local blockchain information to the attached screen. The code for the prototype is available in an open source repository~\footnote{\url{https://gitlab.uzh.ch/mdrezuanul.haque/blockchain-consensus-visualization-on-raspberry-pi}}.

% Visualization component is a separate local server running on each RPi and displaying visual information on the current state of each local ledger - the last two blocks in a color-coded manner. The visualization is designed so that it's easy to see whether a network is in consensus or not.

\vspace{0.25cm}
\vspace{0.25cm}

\subsection{Visualization of consensus}
\begin{figure}[ht]
  \centering
  \begin{minipage}[b]{0.49\textwidth}
    \includegraphics[width=\textwidth]{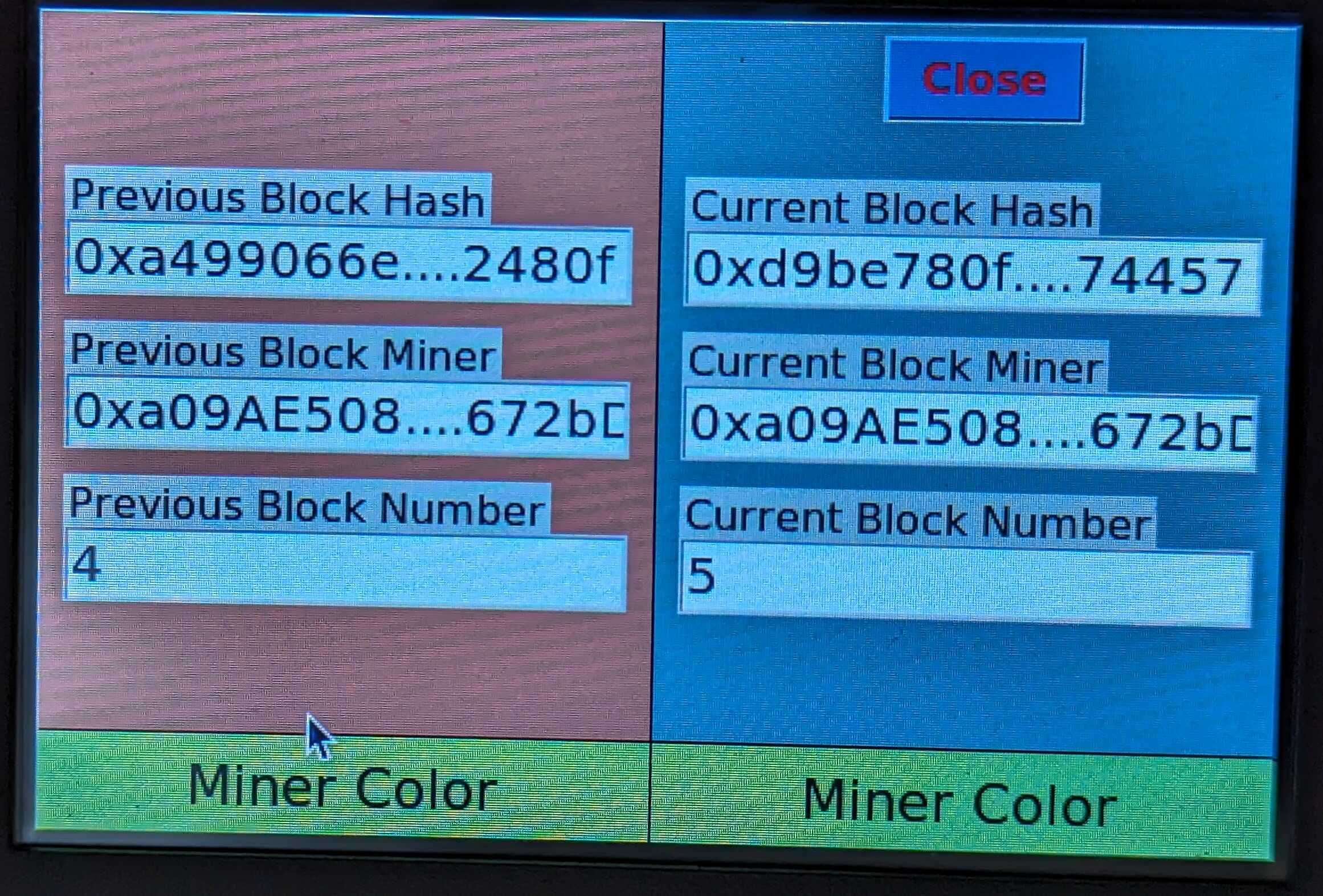}
  \end{minipage}
  \hfill
  \begin{minipage}[b]{0.49\textwidth}
    \includegraphics[width=\textwidth]{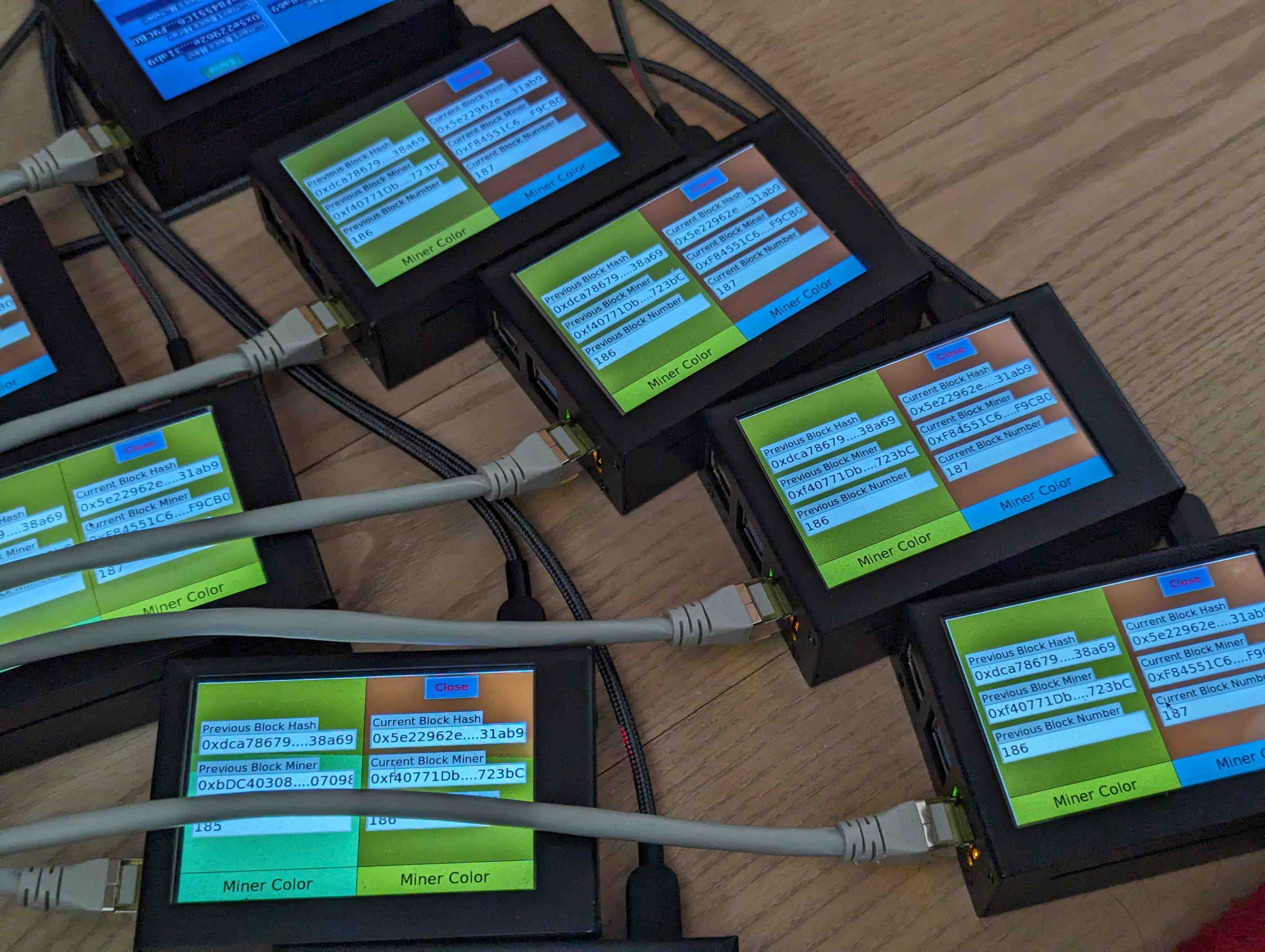}
  \end{minipage}
  \caption{Visualization shown on a single Raspberry Pi screen showing the last two blocks in a local blockchain (left) along with some additional information. The goal of such a simple visualization is to easily see whether current blockchain network is in consensus or not (right).}
  \label{fig:visualization}
\end{figure}

Each Raspberry Pi has an LCD screen attached which shows the current state of the local blockchain - specifically the last two blocks with color coding derived from their hashes so that they are visually distinct. \cref{fig:visualization} shows how the visualization looks like in a typical experimental run. Showing just the last two blocks allows one to easily observe whether current blockchain network is in consensus or not. Also, whenever a new block is mined in the network its propagation through the network should be visible on the screens.

\vspace{0.25cm}
\vspace{0.25cm}

\subsection{Experimental setup}

Through the master server it is possible to adjust several parameters in the blockchain, network and the RPi configuration. The two most important ones that we will describe in this section are the mining \emph{difficulty} and the \emph{topology} of the P2P network~\cite{Grundmann2022}. 

The difficulty parameter is set through a custom genesis block used for blockchain network initialization. It is inversely proportional to the number of block hash calculations required to generate a valid block - for example, difficulty of 20000 means there is approximately 1 in 20000 chance of getting the valid block hash. We use it to set the block generation rate of around 0.75 to 1 second. This generation rate is faster than in the Ethereum mainnet (where a block is being generated every 13 to 20 seconds)~\cite{Buterin2015} which allows us faster experiments, but still high enough so that the natural network environment and protocol communication latency does not affect the results of our experiments. As all of our RPi units have equal computational power it is straightforward to calculate the necessary difficulty parameter for a fixed number of RPi units in order to achieve the desired block generation rate. In our case we used 9 RPi units and the desired block production rate was 0.75 seconds, therefore the required difficulty parameter was calculated and set to 215000, enabling the desired block generation rate in the experimental setup.
%if difficulty is same for all pis, why would latency be an issue

\begin{figure}[ht]
\centering
\includegraphics[width=5in]{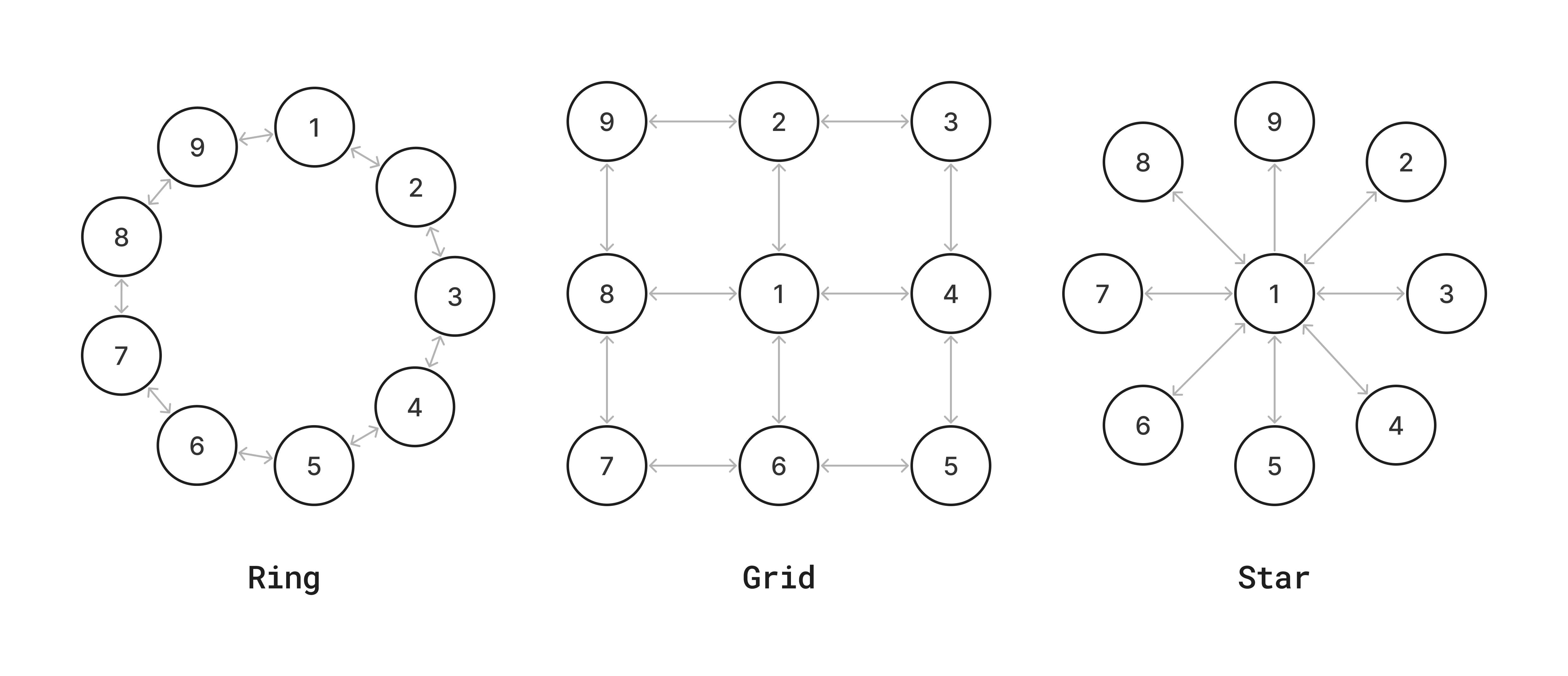}
\caption{P2P topologies used for the experiments with the 9 Raspberry Pi's.}
\label{fig:topologies}
\end{figure}

We define several P2P topologies for our blockchain network - namely the \emph{ring}, the \emph{star} and the \emph{grid} topology. P2P connections between RPi nodes are managed in software through setting the Ethereum client peer connection list - all RPi nodes have preconfigured IP addresses on a local Wi-Fi router that are persistent through the blockchain network initializations. The topologies selected for our experiments are designed to accommodate exactly 9 nodes, making the total hashrate in the network constant throughout the experiments. However, the described system can accommodate any number of nodes and interconnections between them. \cref{fig:topologies} show the three topology types.

\subsection{Consensus quality metrics}
We use several consensus quality metrics in order to evaluate the influence of experimental parameters on the quality of consensus. The \emph{mainchain rate}~\cite{Kraner2023} $\mu$ is defined as the total number of blocks included in the mainchain (the canonical chain of the network), denoted by $|M|$, divided by the total number of produced blocks, denoted as $|B|$:
    \begin{equation}
        \mu = \frac{|M|}{|B|} 
    \end{equation} 
The value of $\mu \in (0;1]$ and the 100\% efficient consensus is achieved when $\mu \rightarrow 1$. This situation is only achieved when every block produced in the network ends up in the mainchain. However, this s never achieved in practice considering that miners are competing for the blockspace and therefore some blocks are never included in the mainchain~\cite{Nakamoto2008}.

The \emph{branching ratio}~\cite{Kraner2023} $F$ quantifies the number of forks in the network, and is defined as:
\begin{equation}
    F = \frac{1}{|M|} \sum_{b \in M} \sum_{c \in \Theta} \delta(p(b), p(c))
\end{equation}
where $p(b)$ is the block parent of block $b$ and $\delta$ is Kronecker delta: $\delta(i; j) = 1$ if  $ i = j$ and $0$ otherwise. The value of $F \in [0; +\infty)$ and a 100\% efficient consensus is achieved when $F = 0$, meaning that there were no accidental forks in the network. 

% @online {
%     blockchainFork,
%     author = "",
%     title = "Blockchain Forks",
%     url = "https://www.geeksforgeeks.org/blockchain-forks/",
%     addendum = "Last accessed: 2024-01-11"
% }

The \emph{contribution ratio} is defined as the  total number of blocks $|N|$ produced by the specified miner node and accepted into the mainchain, divided by the total number of blocks $|M|$ included in the mainchain:
\begin{equation}
        C = \frac{|N|}{|M|} 
    \end{equation}
The value of $C \in [0; 1]$ and it should follow that $N \rightarrow \frac{|M|}{n}$, where n is the number of the miner nodes in the network, if all miner nodes are contributing equally to the mainchain.

% If we consider equal hash rate (computing power) (\cite{hashrate})  of all miner nodes in the network and ideal network conditions, network should provide an equal opportunity to produce a valid block, that will be included in the mainchain. Accordingly .

% @online {
%     hashrate,
%     author = "Hertig, Alyssa and Leech, Ollie",
%     title = "What Does Hashrate Mean and Why Does It Matter?",
%     url = "https://www.coindesk.com/tech/2021/02/05/what-does-hashrate-mean-and-why-does-it-matter/",
%     addendum = "Last accessed: 2024-01-11",
%     year = {2021}
% }

The \emph{initial consensus} is the number $I$ of the mainchain blocks required for the specific node to establish initial consensus with the network by joining the mainchain and downloading chain information to be able to start actively influence the network by producing blocks. Because in our experimental setup configuration each network run starts from the genesis block (block 0), this entails $|I| \in [1;+\infty)$.

\section{Results}

\begin{figure}[ht]
  \centering
  \begin{minipage}[b]{0.49\textwidth}
    \includegraphics[width=\textwidth]{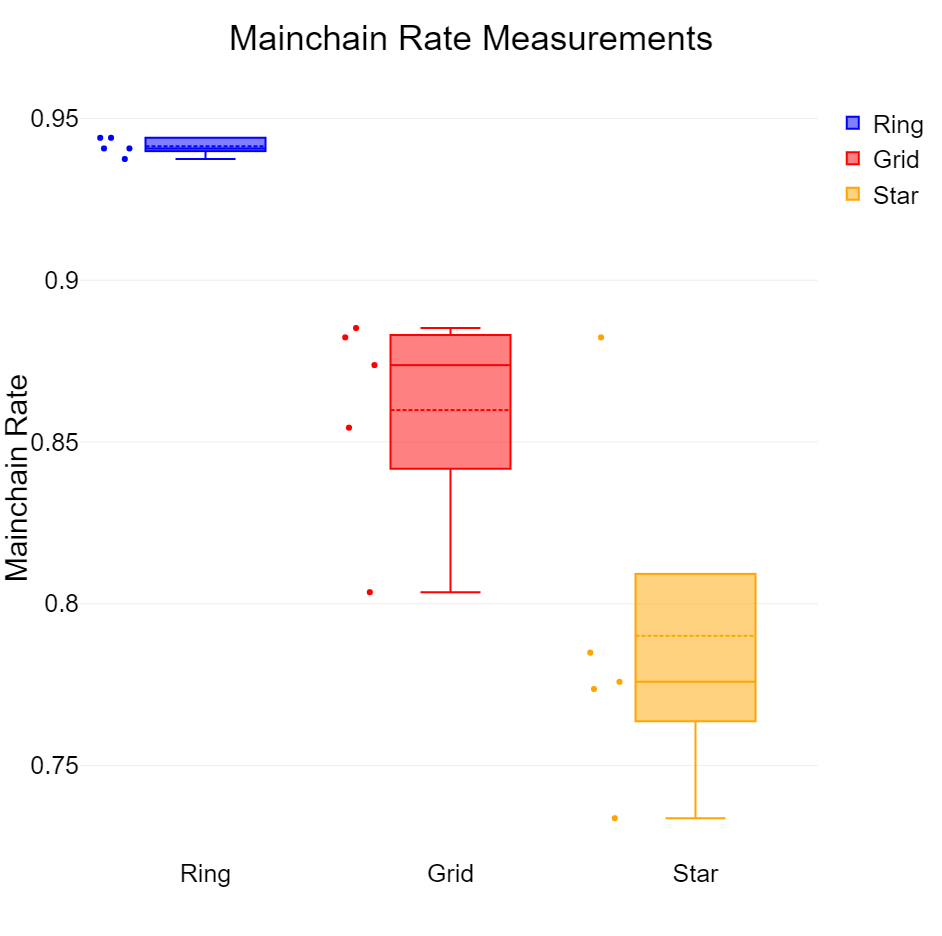}
  \end{minipage}
  \hfill
  \begin{minipage}[b]{0.49\textwidth}
    \includegraphics[width=\textwidth]{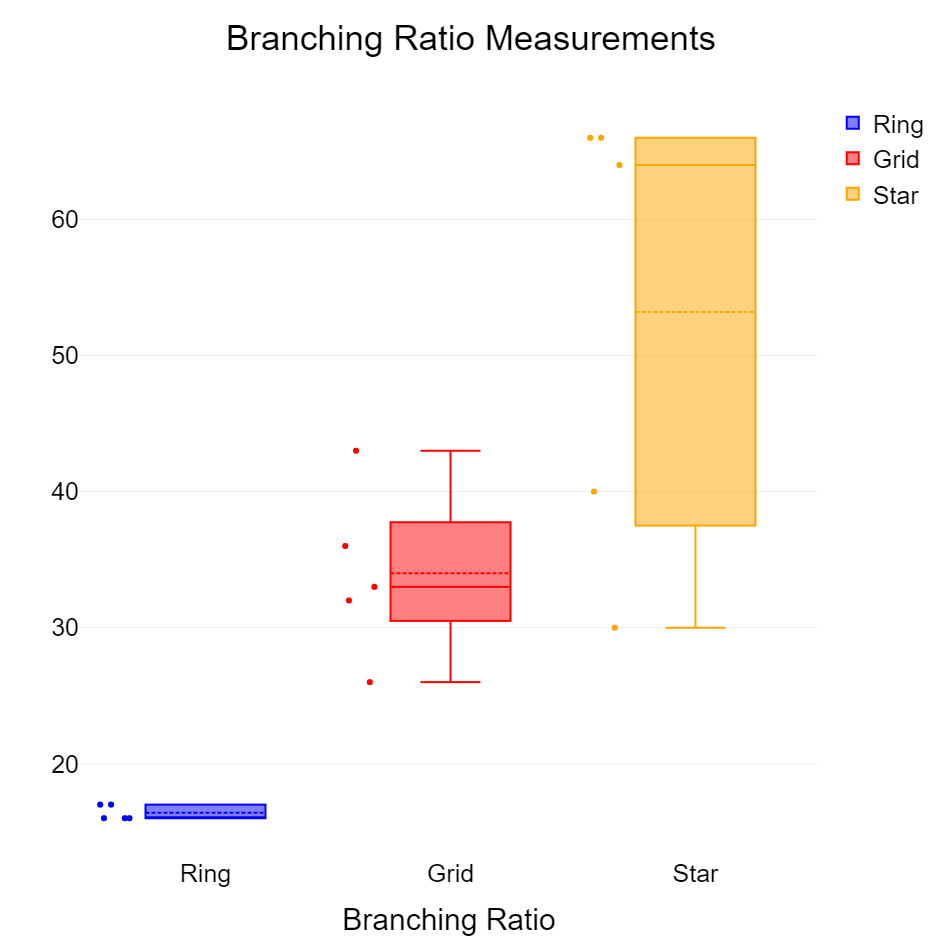}
  \end{minipage}
  \caption{Mainchain rate (left) and branching ratio (right) for the three topologies.}
  \label{fig:results-mainchain-branching}
\end{figure}

\cref{fig:results-mainchain-branching} shows the results of the mainchain rate and the branching ratio measurements on the three topologies. Mainchain rate (higher is better) of the ring topology is significantly higher than in the grid topology, which is in turn higher than in the star topology. The results are counter-intuitive at first because in the ring topology the average shortest distance between nodes is the highest, while in the star topology it is the shortest, implying that the star topology should exhibit the fastest block propagation speed and therefore reach consensus in shortest amount of time, reducing the number of discarded blocks which increases the mainchain rate. However, we have to take into account the additional overhead of managing multiple peer connections and relaying information between them. In the ring topology all communication has to go through the central node, which therefore acts as a bottleneck in consensus. We should note that how node handles peer connections depends also on the specific implementation of the blockchain client. In our experiments we used geth client, which is one of the most popular execution clients for Ethereum, however other clients could exhibit different behavior.

Similarly, the branching ratio (lower is better) in \cref{fig:results-mainchain-branching} indicates that overhead of managing multiple peer connections in the star topology is indeed the reason for its highest branching ratio. Additional observation (not visible in the graphs of \cref{fig:results-mainchain-branching}) is that in the experiments with the highest branching ratio there are certain number of blocks which were never considered by the network for inclusion to the mainchain nor attached to the accidental forks in the network (uncle blocks), indicating clear issues with the consensus.

\vspace{0.25cm}
\vspace{0.25cm}

\begin{figure}[ht]
  \centering
    \includegraphics[width=0.40\textwidth]{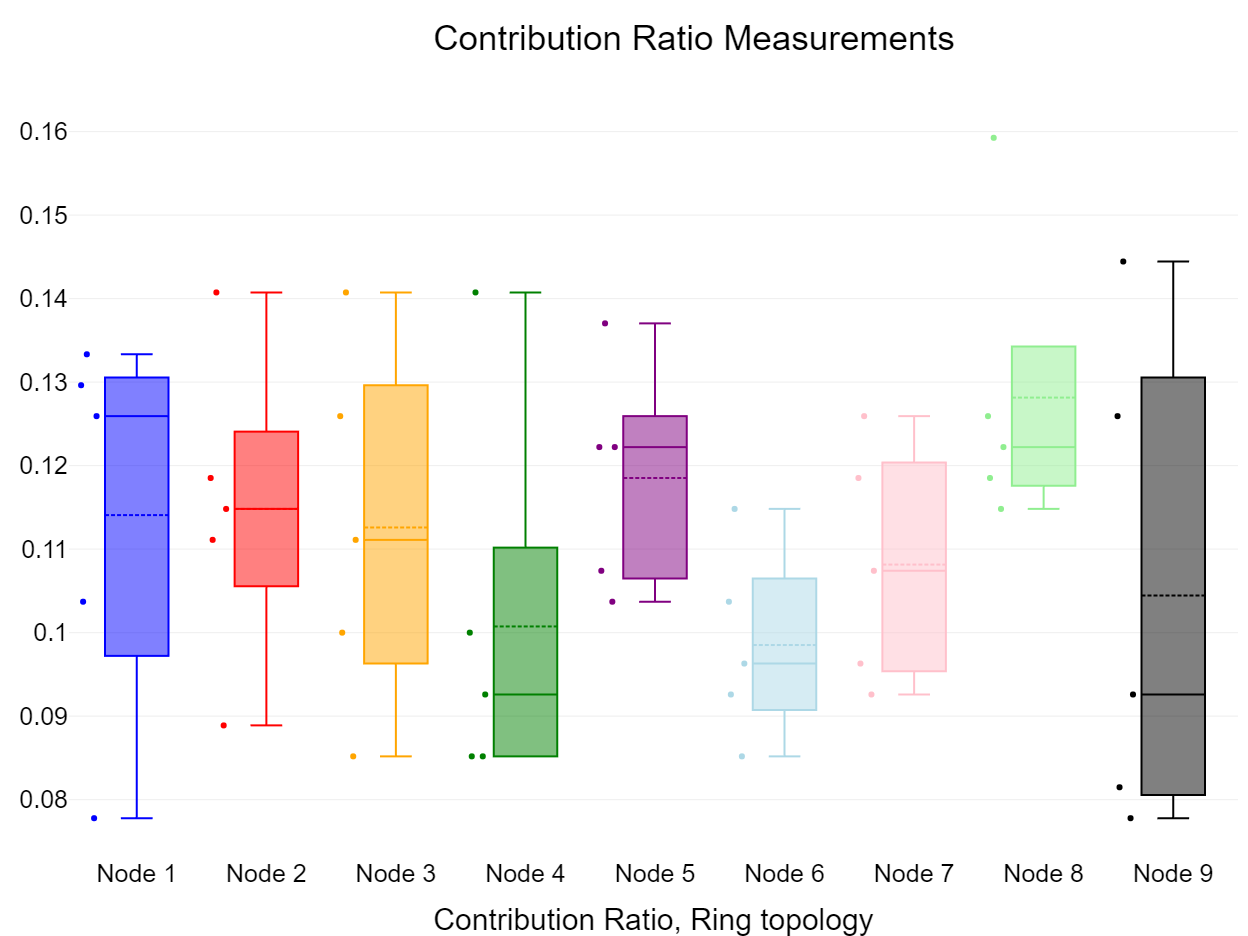}
    \includegraphics[width=0.40\textwidth]{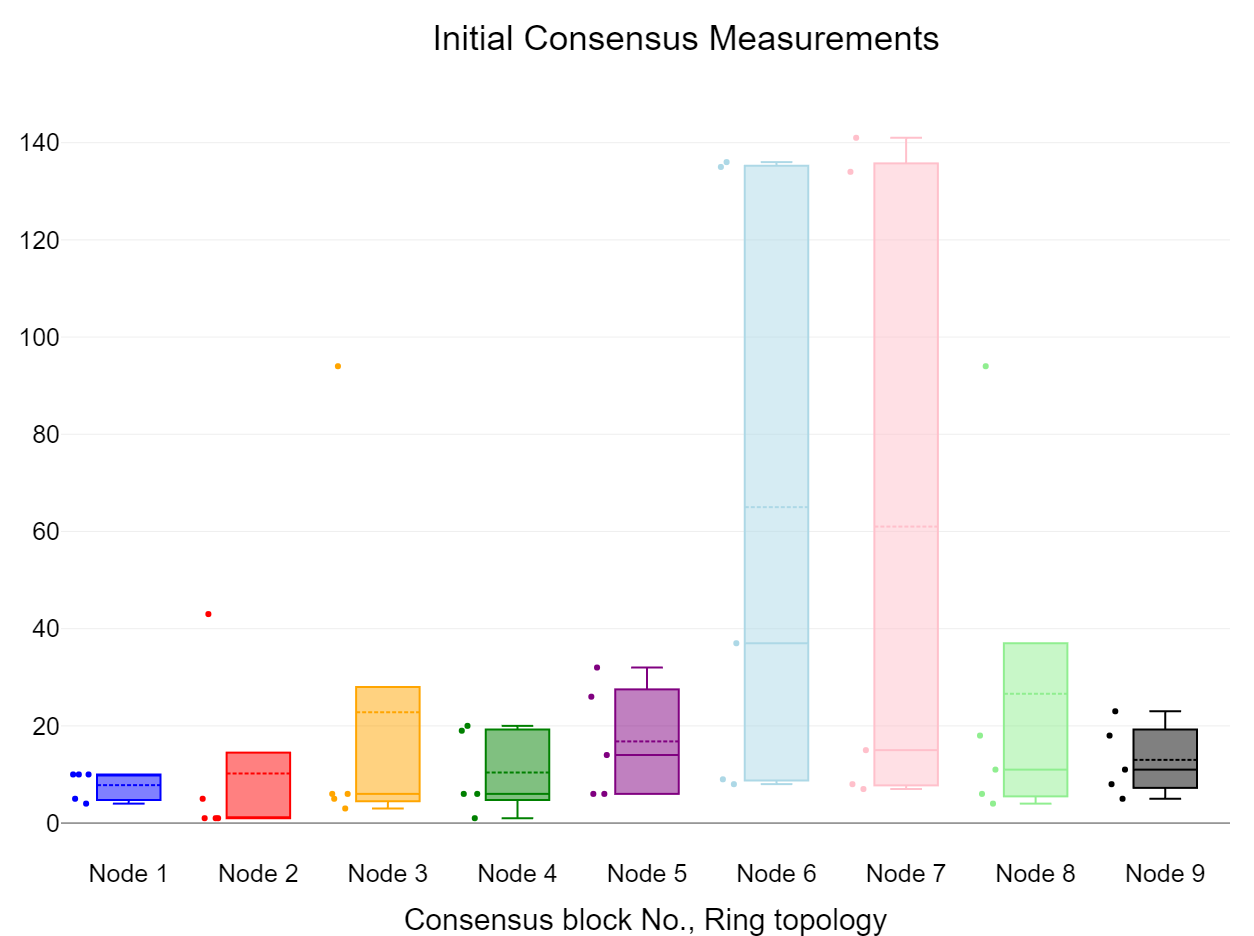}
    \includegraphics[width=0.18\textwidth]{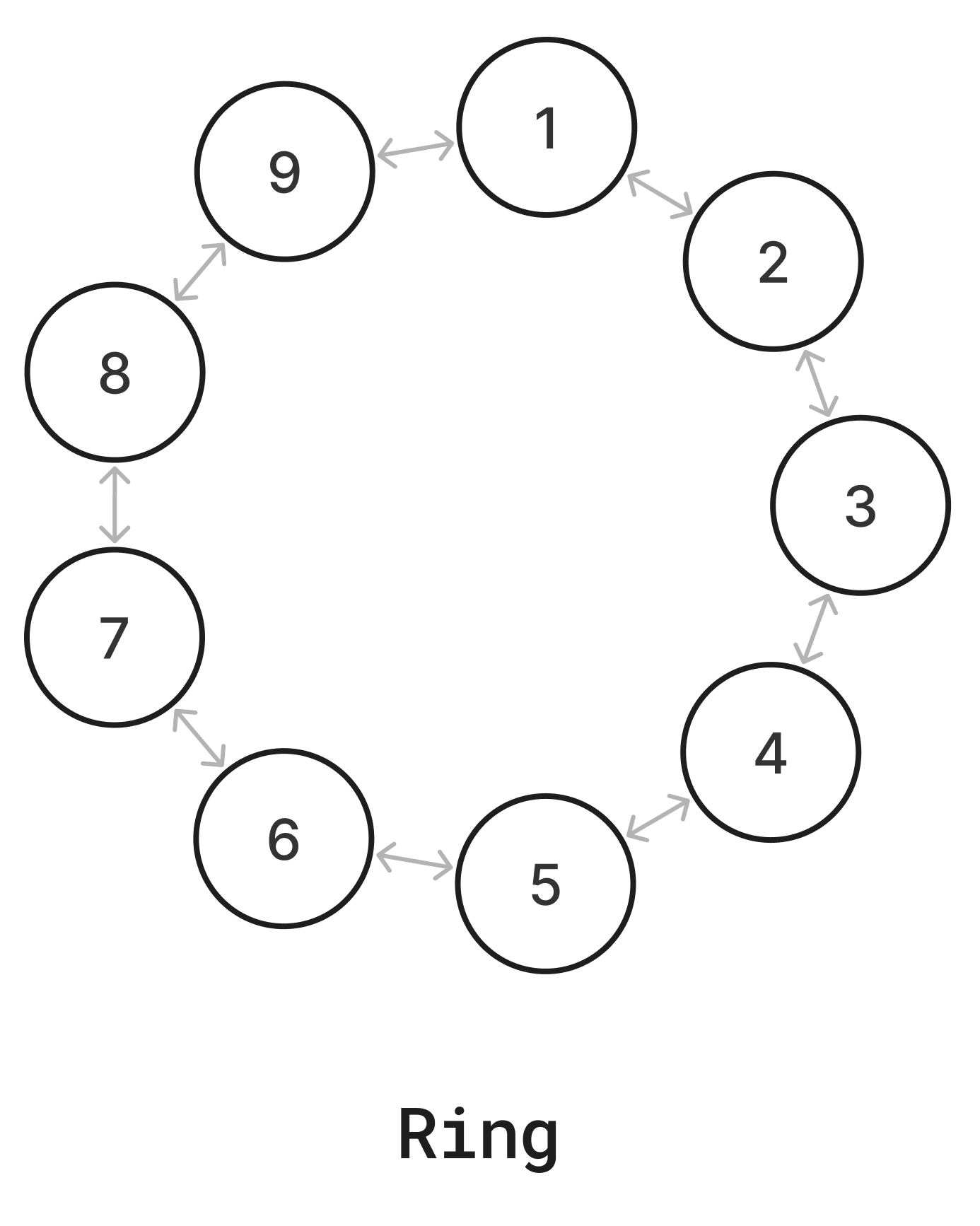}
    \includegraphics[width=0.40\textwidth]{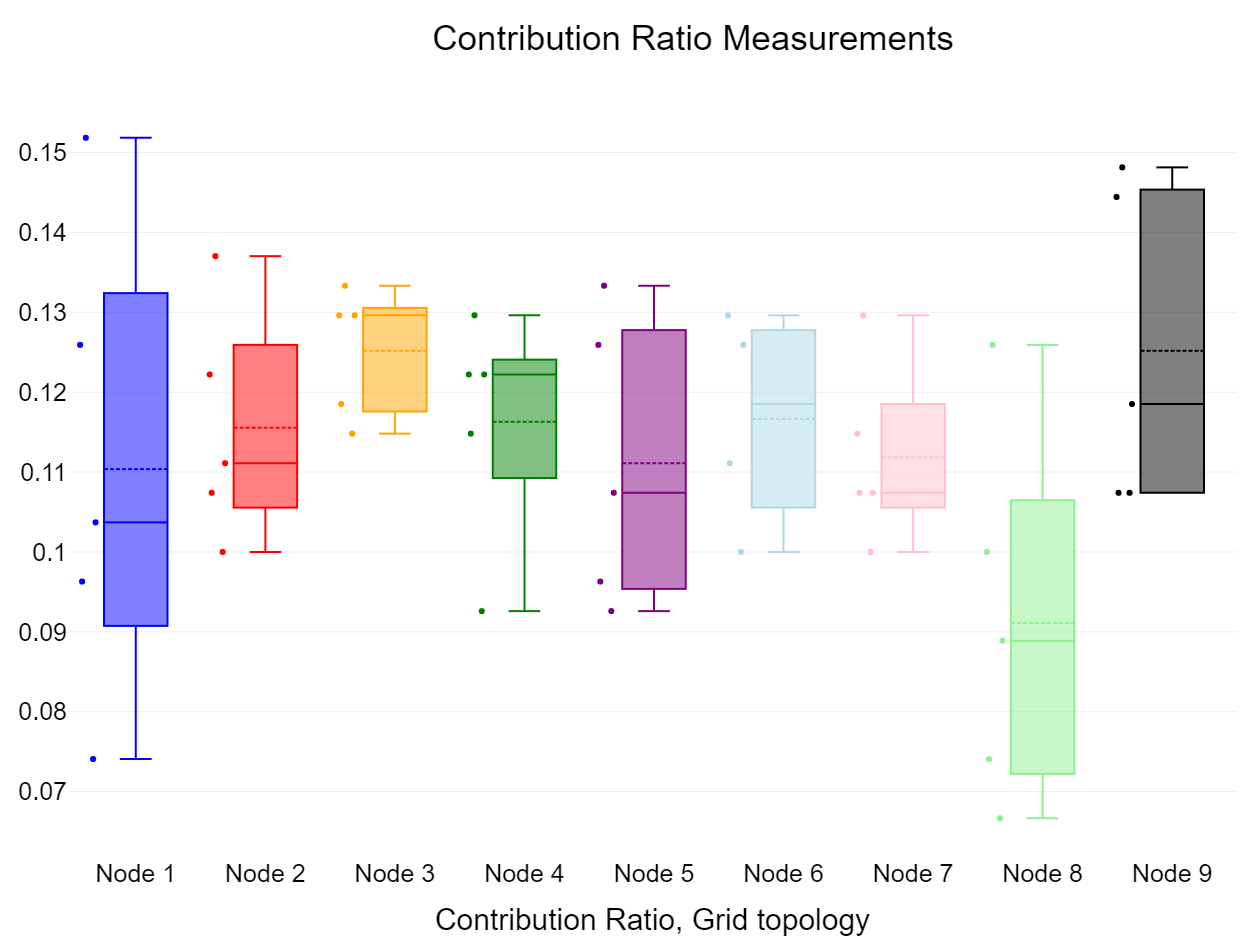}
    \includegraphics[width=0.40\textwidth]{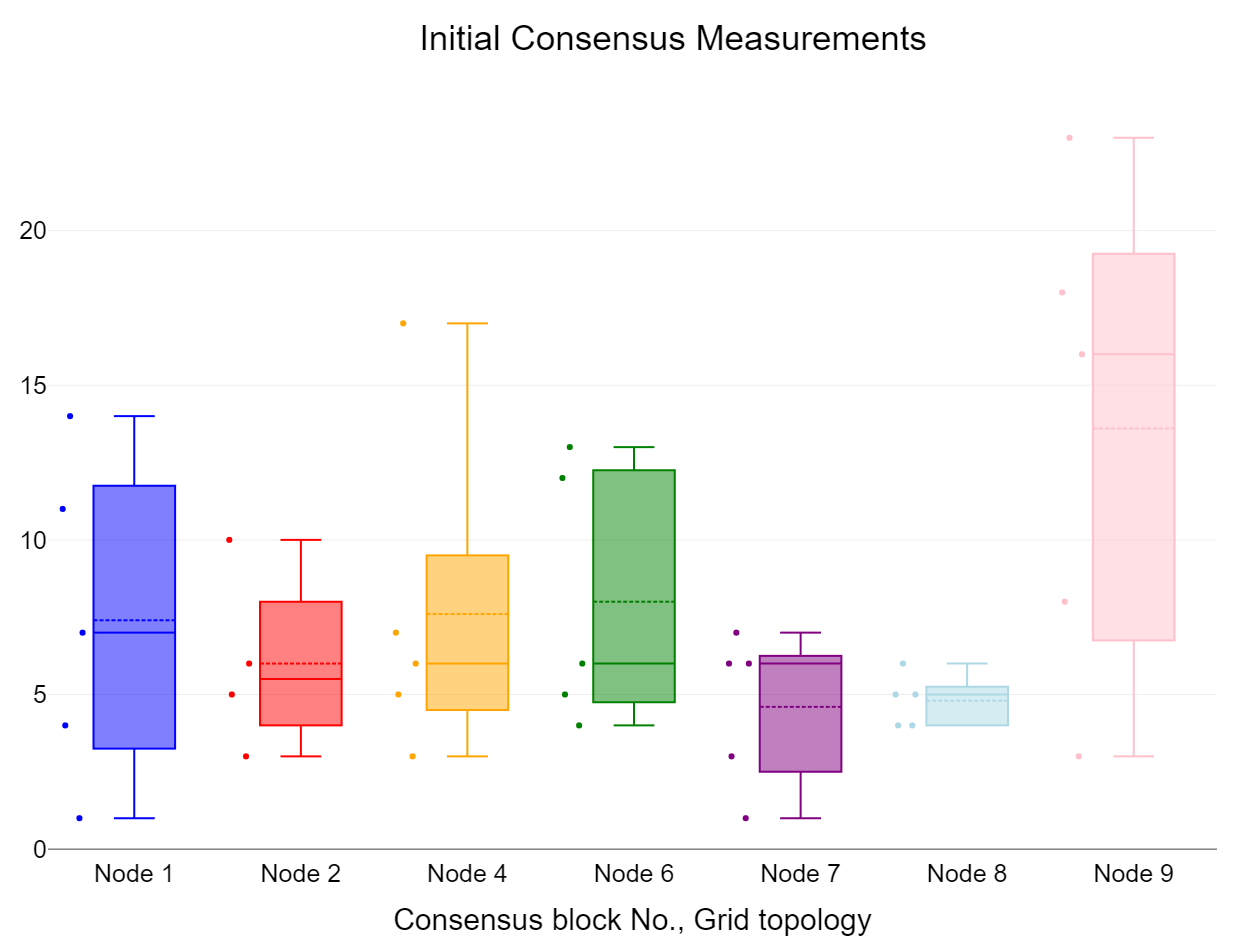}
    \includegraphics[width=0.18\textwidth]{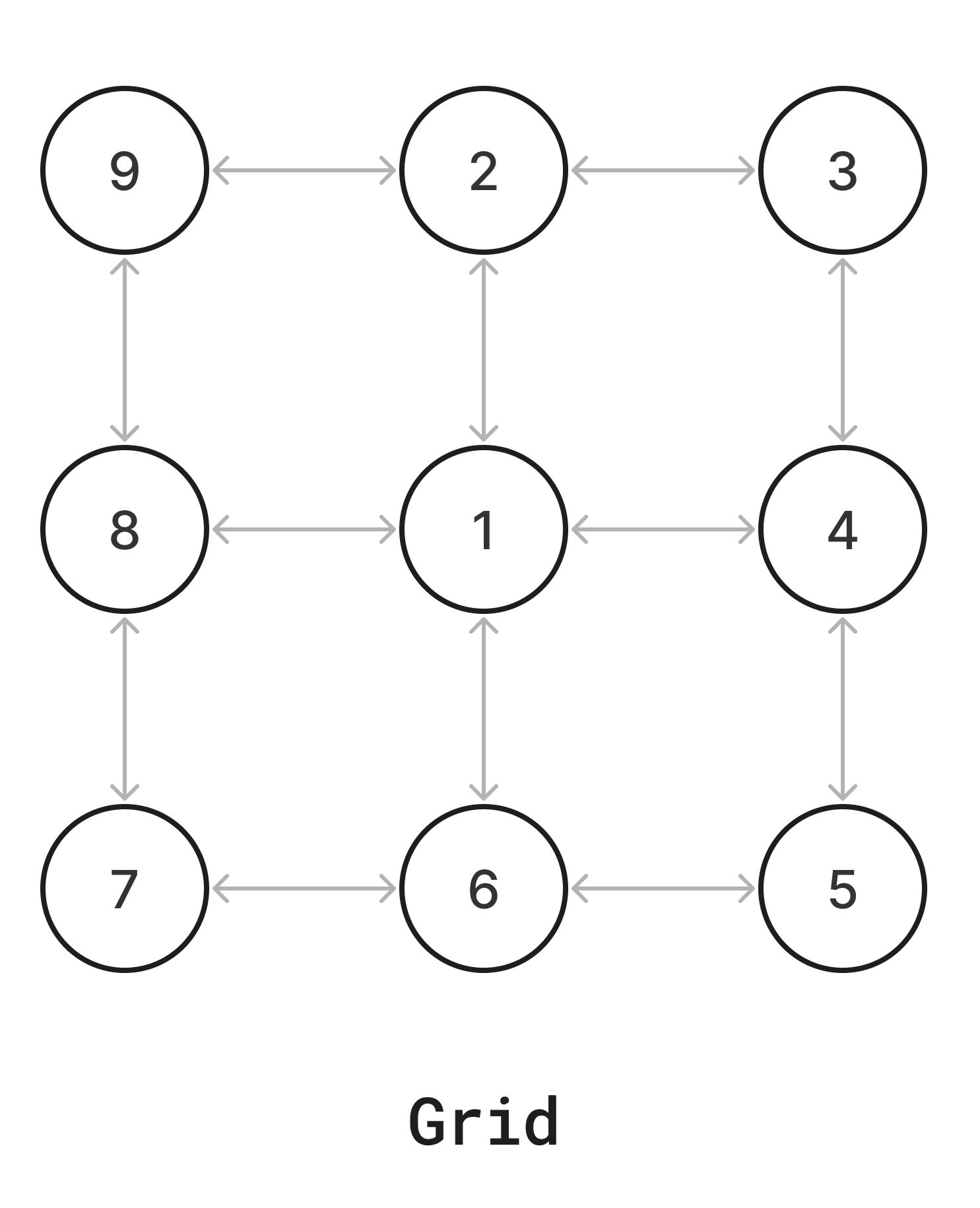}
    \includegraphics[width=0.40\textwidth]{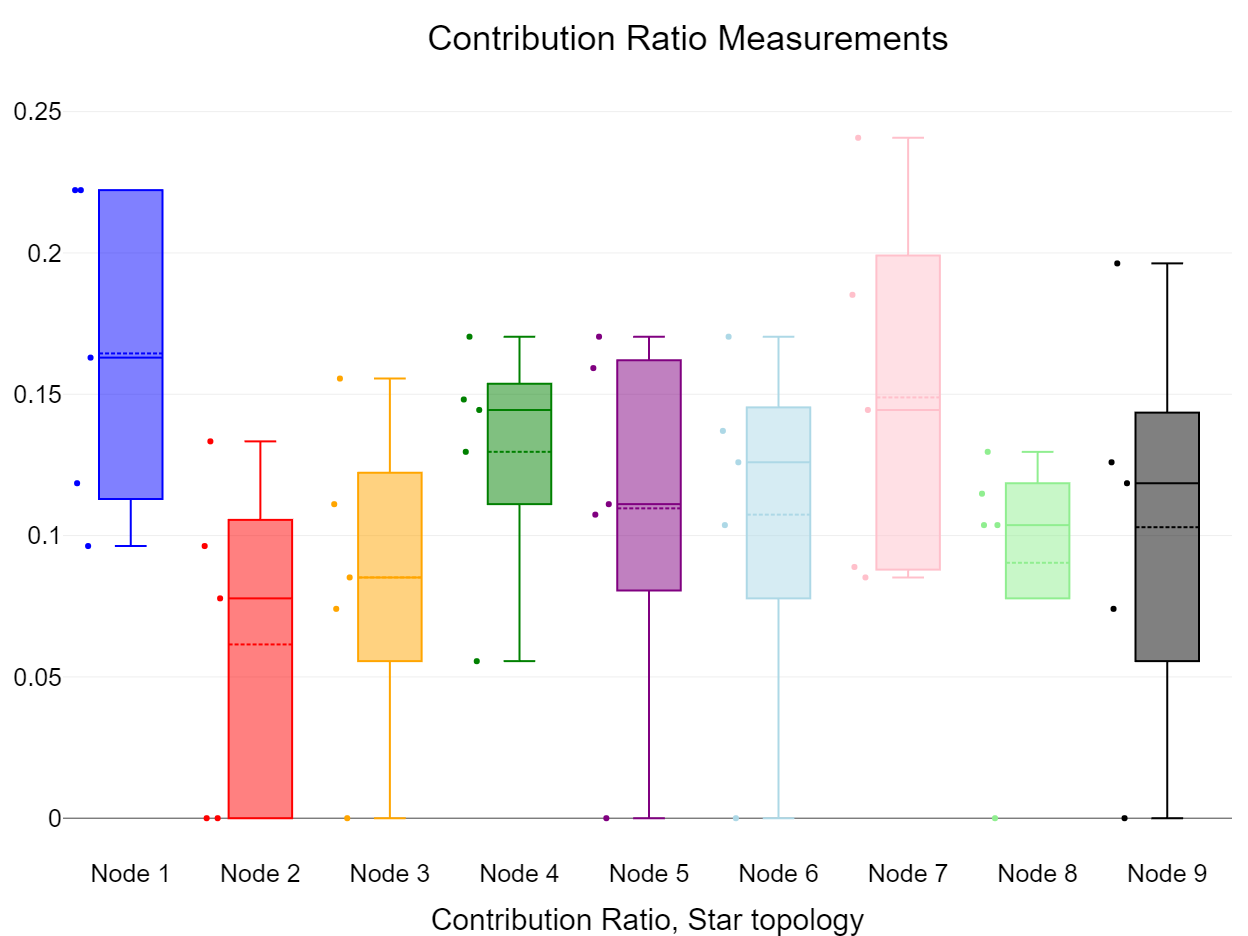}
    \includegraphics[width=0.40\textwidth]{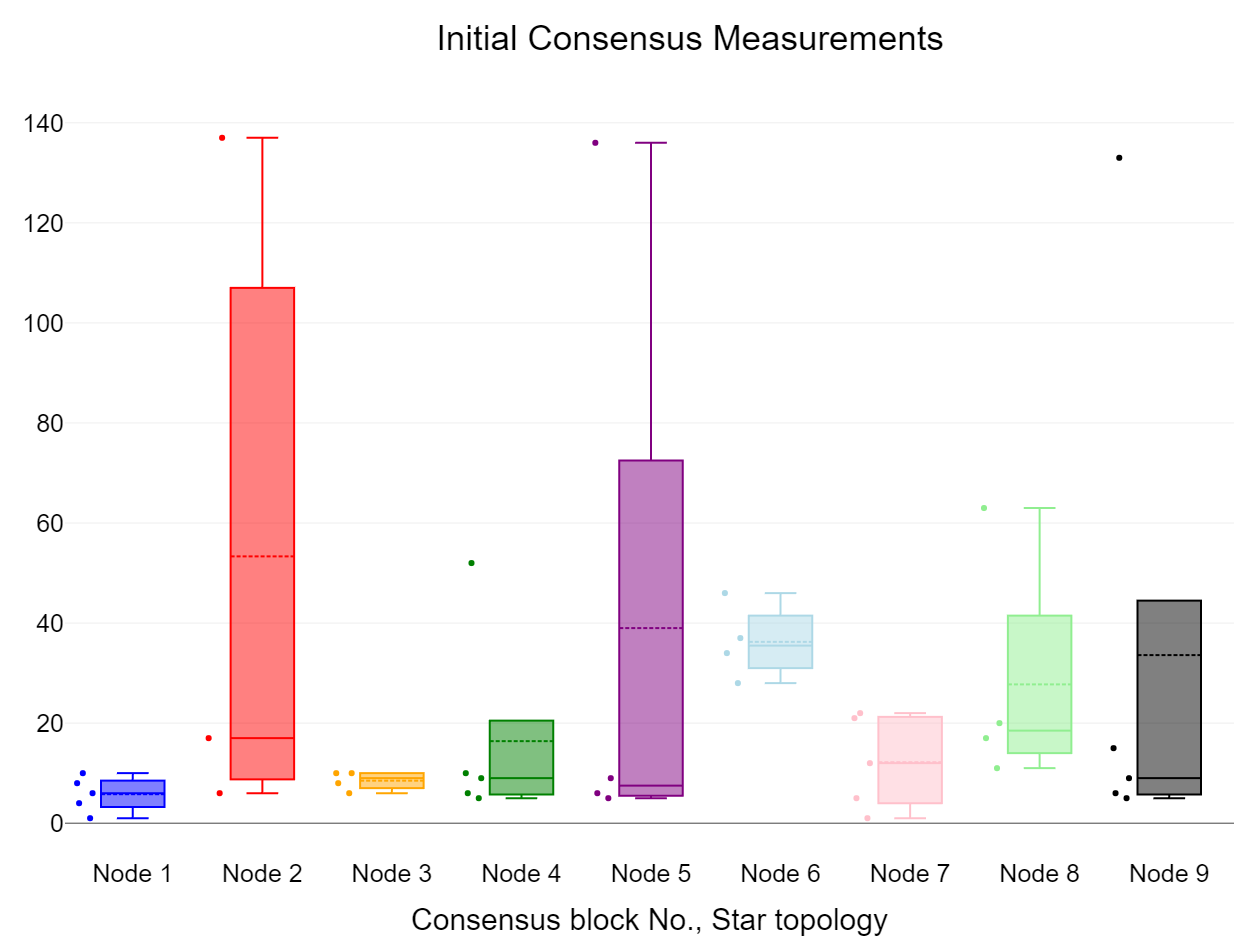}
    \includegraphics[width=0.18\textwidth]{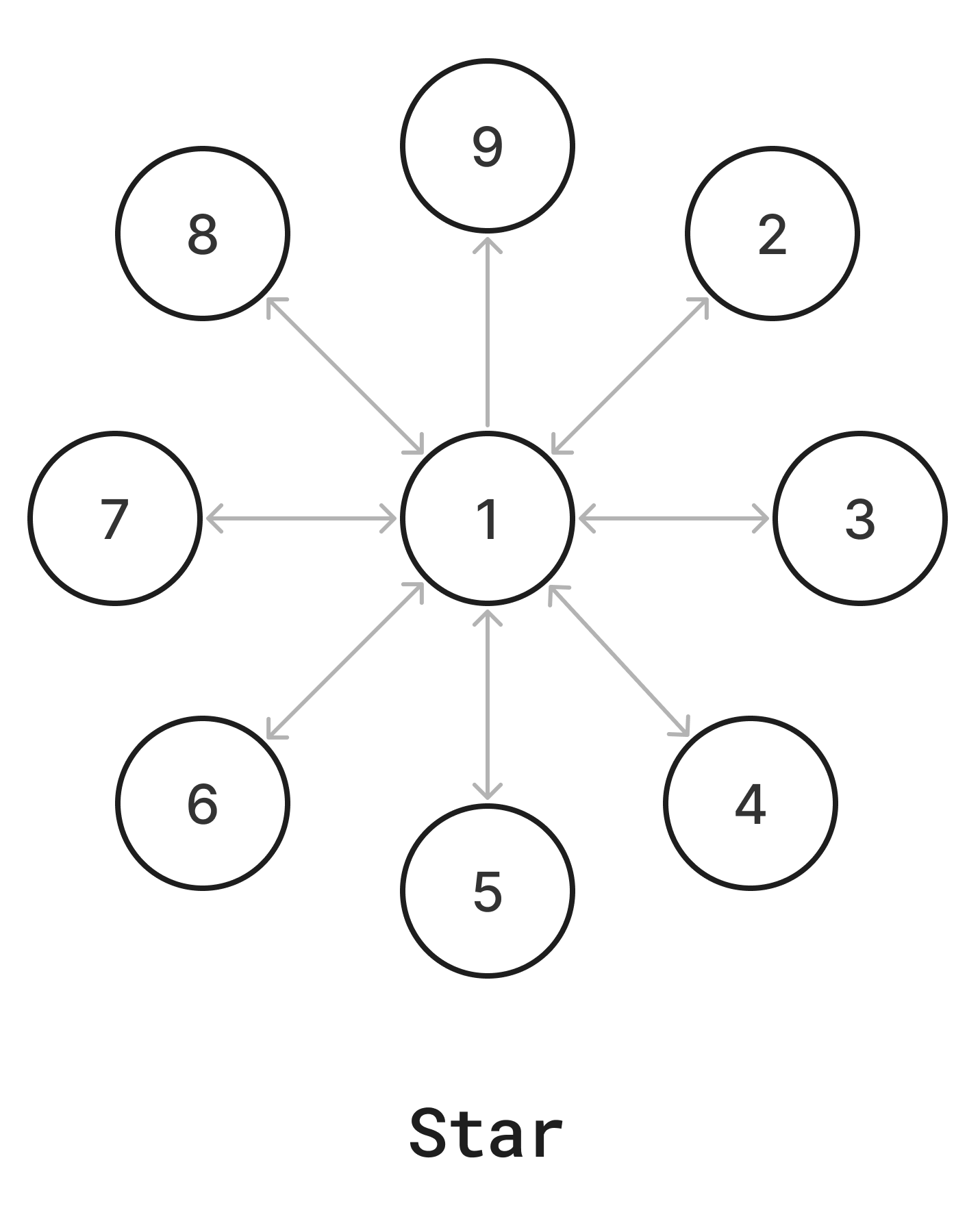}
  \caption{Contribution ratio and the initial consensus for the three topologies.}
  \label{fig:results-contribution}
\end{figure}

In the ideal blockchain network consisting of nodes with equal mining power and and fast network communication, the contribution ratio of each node is expected be equal. In our case where there are 9 RPi nodes the contribution ratio of each node should be $1/9=0.11$. \cref{fig:results-contribution} shows the contribution ratios for the three topologies in our experiments. There is certain variation in the measurements, with some outliers, most notably with the star topology where central node dominates the consensus (which is expected due to its central position in the topology~\cite{Fadda2022}) while, at the same time, several nodes never contributed to the mainchain. This is most probably due to unfinished initial synchronization of the blockchain (visible in the middle graph showing the initial consensus measurement). This implies that restrictive peer connections (like in the case of star topology, where the whole network is connected through a single node) can severely impact the quality of the consensus.

\cref{fig:results-contribution} also shows that, on average, initial consensus usually takes less than 25 blocks. However, there are cases where some nodes struggle with initial synchronization and are able to join the network only after 100+ blocks. The issue here is that nodes that are joining the network are depending on one specific node to download the current chain, however if that ``parent'' node restructures its own chain then the initial synchronization will fail before node decides to receive the new state of the network. The solution for this would be to manually reset the synchronization as soon as such situation is detected. For now we just removed from the plots all cases where nodes are unable to achieve initial consensus, since they do not have a numerical value.

\vspace{0.25cm}
\vspace{0.25cm}

\section{Discussion}

Experiments described in the last section demonstrate that our blockchain-in-a-box prototype is capable enough to conduct consensus quality measurements in a completely isolated environment, not even using an Internet connection, yet running a fully functional version of the Ethereum PoW blockchain network. Running a blockchain network through a hardware setup, as opposed to a software simulation, results in a much more realistic behaviour. For example, measurements in all of the experiments above are significantly influenced by the communication overhead between nodes - a fact that node has finite computational capabilities and that, on top of consensus maintenance and mining, it has to relay messages between nodes. This makes network P2P topology a crucial factor in achieving efficient consensus, as measured by the mainchain rate and the branching ratio. We also have to account for the implementation logic of the specific Ethereum node and mining client (in our case geth) which was not created with certain usage scenarios in mind - for example, to have a very limited connectivity to a single node such as in the star topology that we use. These kinds of insights would be hard to reproduce with a pure software simulation.

\section{Conclusion}

This paper describes a fully functional blockchain-in-a-box prototype consisting of multiple Raspberry Pi nodes running a private version of the Ethereum PoW protocol, together with a master server and a web-based interface allowing users to bootstrap and orchestrate their operation in order to conduct experiments under various conditions. Each RPi unit has a screen attached which shows the content of its local blockchain, making it easy to observe whether nodes are in consensus or not. Developed prototype does not depend on the external Internet capability so it is appropriate to be used as a demonstration of a fully functioning blockchain system in a classroom or a workshop setting. In addition to educational value, the prototype allows one to conduct experiments on a fully functioning blockchain network that would be hard or impossible to conduct on nodes running Ethereum mainnet nodes, for example measuring consensus quality under various experimental conditions such as P2P topology and similar.

%define the following sections to hide their Section Number (Notes Style)
\ledgernotes

\section{Acknowledgement}
This work was funded in part by the grant from DLT Science Foundation.

\section{Author Contributions}
AI, SHA, LA and MRH developed hardware and software components of the prototype. AI performed the experiments. MP, AI, MC and CJT designed the research methodology. MP, MC and CJT supervised the execution of the experiments. All authors contributed equally to manuscript preparation.

% \section{Conflict of Interest}
% \todo{This section may be omitted if no conflict of interest exists.}

%AUTHOR: comment out if using thebibliography
% \theendnotes

%AUTHOR: comment out, this is used to make sure the Creative Commons License
%image fits on page
\newpage

% %define the following sections to have the Appendix Style
% \appendix

% \section{Calculations}
% Relegate to an appendix material that would distract from the flow of the article, but that is required to rigorously prove claims made or concepts introduced in the paper’s body.

% AUTHOR: this is an alternative to using Endnotes
\bibliographystyle{nature}

%add the Creative Commons license footer to the last page
\thispagestyle{pagelast}

\end{document}